\documentstyle[aps,amssymb,12pt]{revtex}

\begin{document}
\author{S. Bruce\thanks{%
Electronic address: sbruce@udec.cl; Fax: (56-41) 22 4520.} and J. Diaz-Valdes}
\address{Physics Department, University of Concepcion, P.O. Box 160-C,\\
Concepcion,Chile}
\title{An approximation to the Lamb shift and hyperfine splitting as nonlinear
effective Coulomb-like interactions in the Dirac equation}
\date{May, 1999}
\maketitle

\begin{abstract}
The Dirac equation for the Coulomb problem is restated by incorporating a 
{\it nonlinear effective} interaction into the Dirac Hamiltonian: one keeps
the $1/r$ dependence for the Coulomb field, but the coupling constant is
modified by a factor depending on the $n$ ({\it principal quantum number})
power of the mean value of the Hamiltonian$.$ In this simple context we
study the Lamb shift and the hyperfine splitting of the $s$-levels of
hydrogenic atoms. We discuss to what extent the corresponding calculations
fit the energy splittings to the appropriate order in the {\it fine} {\it %
structure} constant.

PACS number(s): 03.65.P, 32.10.F, 31.30.J
\end{abstract}

\bigskip

\section{Introductory Background}

In 1947, Lamb and Retherford \cite{LA,LA1,LUN} observed that the $2s_{1/2}$
and the $2p_{1/2}$ energy levels of the hydrogen atom were split: the $%
2p_{1/2}$ energy level was depressed more than 1000 MHz below the $2s_{1/2}$
energy level. The original theory of a Dirac electron in a classical Coulomb
potential predicted that the energy levels of the hydrogen atom should
depend only on the {\it principal quantum number} $n$ and the total spin $j$%
, so these two levels should be degenerate.

The calculation of the Lamb-shift is rather intricate, because one is
dealing with the hydrogen atom ($Z=1$) as a bound-state problem, and also
because we must sum over all radiative corrections to the electron
interacting with a Coulomb potential that modify the na\"{i}ve $\Psi
^{\dagger }A_{0}\Psi $ vertex. These corrections include the vertex
correction, the anomalous magnetic moment, the self-energy of the electron,
the vacuum polarization graph, and even infrared divergences.

The original nonrelativistic bound-state calculation of Bethe \cite{BE},
which ignored many of these subtle higher-order corrections, could account
for about 1000 MHz of the Lamb-shift, but only a fully relativistic quantum
treatment could calculate the rest of the difference. To begin the
discussion, we first see that the vacuum polarization term can be attached
to the photon line, changing the photon propagator. This, of course,
translates into a shift in the {\it effective coupling} of an electron to
the Coulomb potential \cite{UE}. We know from ordinary nonrelativistic
quantum mechanics that, by taking matrix elements of this modified potential
between Coulomb wave functions, we can calculate the first-order correction
to the energy levels of the one-electron atom due to the vacuum polarization
graph.

This discussion can be generalized to include the other corrections to the
calculation of the Lamb-shift. The method is the same: corrections to the
vertex function $\Psi ^{\dagger }\gamma _{\mu }\Psi $, take the zeroth
component, and then take the low-energy limit. If we add the various
contributions to the vertex correction, we find the well-known formula: 
\begin{equation}
\Delta E_{n\kappa }^{{\rm (Lamb)}}=\frac{4mc^{2}}{3\pi n^{3}}\alpha \left(
Z\alpha \right) ^{4}\left( L_{n\kappa }+\left( \frac{19}{30}-2\ln \left(
Z\alpha \right) \right) \delta _{\kappa ,-1}+\frac{3}{8}\frac{1}{\kappa
\left( 2\left| \kappa \right| -1\right) }\left( 1-\delta _{\kappa
,-1}\right) \right) ,  \label{e}
\end{equation}
to order $\alpha \left( Z\alpha \right) ^{4}\ln \left( Z\alpha \right) ,$
where $m$ is the mass of the electron and $\alpha =e^{2}/\hbar c$ is the 
{\it fine structure constant}. The term 
\begin{equation}
L_{n\kappa }\equiv \frac{n^{3}}{2m^{2}c^{2}\left( Z\alpha \right) ^{4}}%
\sum_{n^{\prime }\kappa ^{\prime }}\left\{ \left| <\Psi _{n^{\prime }\kappa
^{\prime }}^{{\rm (NR)}}|\widehat{p}|\Psi _{n\kappa }^{{\rm (NR)}}>\right|
^{2}\times \left( E_{n^{\prime }\kappa ^{\prime }}-E_{n\kappa }\right)
\right.  \label{b}
\end{equation}
\[
\times \left[ 2\ln \left( Z\alpha \right) -\ln \left( \frac{mc^{2}/2}{%
E_{n^{\prime }\kappa ^{\prime }}-E_{n\kappa }}\right) \right] \times \left.
\left( E_{n^{\prime }\kappa ^{\prime }}-E_{n\kappa }\right) \times \left[
2\ln \left( Z\alpha \right) -\ln \left( \frac{mc^{2}/2}{E_{n^{\prime }\kappa
^{\prime }}-E_{n\kappa }}\right) \right] \right\} 
\]
is also known as the {\it Bethe logarithm }which{\it \ }has to be evaluated
numerically. In the above $\kappa =\pm \left( j+1/2\right) $ for $l=$ $j\pm
1/2$, and $\Psi _{n\kappa }^{{\rm (NR)}}$ is a {\it nonrelativistic} atomic
state.

The vertex correction, for $Z=1$ for example, gives us a value of $1010$
MHz. The anomalous magnetic moment of the electron contributes $68$ MHz. The
vacuum polarization graph contributes $-27.1$ MHz. Adding these corrections
together, we find, to the lowest loop level, the Lamb-shift to within $6$
MHz accuracy.

Since then, higher-order corrections have been calculated, so that the
difference between experiment and theory has been reduced to $0.01$ MHz .
Theoretically, the $2s_{1/2}$ level is above the $2p_{1/2}$ energy level by $%
1057.864$ MHz. The experimental result is $1057.862$ within $0.02$ MHz. This
is an excellent indicator of the basic correctness of QED \cite{SA,BER,LAU}.

Another important effect, not contained in the energy levels of the
Dirac-Coulomb problem, arises from the interaction between the magnetic
moment of the nucleus and the magnetic moment of the electron. In the case
of the hydrogen atom, for instance, when we combine the electron spin with
the proton spin, the net result is $S=1$ (triplet) or $S=0$ (singlet), where 
$S$ is the quantum number corresponding to the {\it total spin}. Since the
magnetic interaction is dependent on the relative orientation of the two
magnetic dipole moments, each level of the hydrogen atom characterized by $%
njl,$ is further split into two sublevels corresponding to the two possible
values of $S$ even in the absence of any external magnetic field. This is
known as the {\it hyperfine splitting }\cite{LAU,KL,CRE,SU}.

Using the nonrelativistic wave function $\Psi _{n\kappa }^{{\rm (NR)}}$, we
obtain the energy shift 
\begin{equation}
\Delta E_{n\kappa }^{{\rm (Hyp)}}=\frac{1}{3}g_{p}\left( \frac{m}{M_{p}}%
\right) \frac{mc^{2}}{n^{3}}\left( \delta _{S,1}-3\delta _{S,0}\right)
\delta _{\kappa ,-1}\alpha ^{4},  \label{e2}
\end{equation}
where $g_{p}=2\left( 1+\kappa _{p}\right) =5.58568$ is the $g$-factor of the
proton. Note that the order of magnitude of this splitting is the {\it fine}-%
{\it structure splitting} multiplied by $m/M_{p}$. For the $2s_{1/2}$-state,
the above energy difference corresponds to a radio microwave of $1420$ MHz.
It is an accurately measured quantity: $1420.40575180$ within at least $%
3\times 10^{-8}$ MHz \cite{CR}.

There are other corrections to the Dirac formula. First, we must take into
account the motion of the nucleus since the mass of the nucleus is not
infinite. A major part of this correction can be taken care of if we use the
reduced mass in place of $m$. Second, the finite size of the nucleus,
especially for the $s$ states which are sensitive to small deviations from
Coulomb's law at close distances: in the interesting case of the $2s$ state
of the hydrogen atom, however, we can estimate the energy shift due to this
effect to be only $0.1$ MHz.

The utility of Dirac theory in atomic physics is not limited to light
hydrogen-like atoms. For heavy atoms where $Z\alpha $ is not very small
compared with unity, the relativistic effects must be taken into account
even for understanding the qualitative features of the energy levels.
Although we cannot, in practice, study one-electron ions of heavy atoms, it
is actually possible to check the quantitative predictions of Dirac theory
by looking at the energy levels of the innermost electrons of high $Z$ atoms
which can be inferred experimentally. Similar studies have been carried out
with muonic atoms \cite{CA}.

This paper concerns a simple, and restricted, approach to the study of the
splitting structure of the Dirac-Coulomb energy levels, given in Eqs.(\ref{e}%
) and (\ref{e2}), in the context of relativistic quantum mechanics. This
procedure {\it circumvents} second quantization on both the electromagnetic
and the electron fields. However, it does not pretend to be an alternative
way of reproducing the corresponding detailed and precise modern
calculations by QED. In fact, for the case of the Lamb-shift, we do in fact
make {\it use} of second quantization information (the self-energy of the
electron, the vertex correction, the anomalous magnetic moment, and the
vacuum polarization graphs to order $\alpha \left( Z\alpha \right)
^{4}mc^{2} $) to define the basic coupling constant. In Sec. II, we present
the general approach. It considers the introduction of a {\it nonlinear} 
{\it effective} interaction into the Dirac Hamiltonian. Although the
radiative processes involved in the Lamb-shift will somehow be hidden in the
corresponding effective Hamiltonian, it is illustrative to view them from a
different perspective. The case of the {\it hyperfine structure} is treated
in the same framework, though it does not involve radiative corrections to
the first level, i.e., when only relativistic corrections are taken into
account.

Higher order corrections to the Lamb-shift to order $\alpha \left( Z\alpha
\right) ^{5}mc^{2}$ and to the {\it hyperfine-splitting} are considered in
Sec. III. Finally, Sec. IV contains our conclusions and some open questions.

\section{Effective nonlinear Coulomb-like interaction}

In this paper we want to show that the splittings (\ref{e}) and (\ref{e2})
can be derived from the Dirac-Coulomb problem by incorporating a {\it %
nonlinear} {\it effective} interaction into the Dirac Hamiltonian. To this
end, we shall assume that the interaction does not modify the $\backsim (1/q%
{\bf )}$ ($q\equiv \left| {\bf q}\right| $) law in the Coulomb-like gauge
(central) field. This {\it radial} behavior is one of the few cases where
the Dirac wave equation can be solved analytically. The problem to study is
then the following 
\begin{equation}
H\Psi ({\bf q},t)=\left( -i\hbar c{\bf \alpha \cdot \nabla }_{{\bf q}}{\bf +}%
\beta mc^{2}+eA_{0}^{{\rm (eff)}}\left( q\right) \right) \Psi ({\bf q}%
,t)=i\hbar \frac{\partial }{\partial t}\Psi ({\bf q},t),  \label{h}
\end{equation}
in the standard Dirac representation \cite{IT}. Here the normalization
condition is 
\begin{equation}
\int_{{\Bbb R}^{3}}\Psi ^{\dagger }({\bf q},t)\Psi ({\bf q},t)\ d^{3}q=1,
\end{equation}
and 
\begin{equation}
A_{0}^{{\rm (eff)}}\left( q\right) \equiv \widehat{{\rm g}}_{\nu }\frac{%
Z\left| e\right| }{q}
\end{equation}
is an {\it effective} potential, where $\widehat{{\rm g}}_{\nu }$ is a {\it %
diagonal} constant matrix: It is taken not to be $q$-dependent since we
shall not consider corrections in the vicinity of the nucleus. The
dimensionless elements of this matrix depend on the quantum numbers
represented by $\nu ,$ labeling the stationary states $\Psi _{E_{\nu }}({\bf %
q},t)=\exp \left( -iE_{v}t/\hbar \right) \Psi _{E_{\nu }}({\bf q)}$, whose
concrete structure will be specified, in each case, below.

Before solving the eigenvalue problem associated with (\ref{h}), we recall
that the operators 
\begin{equation}
\widehat{K}\equiv \beta \left( {\bf \Sigma \cdot }\widehat{{\bf L}}+\hbar
\right) ,\qquad \widehat{{\bf J}}{\bf \equiv }\widehat{{\bf L}}{\bf +}\frac{%
\hbar }{2}{\bf \Sigma ,}  \label{c}
\end{equation}
with $\widehat{{\bf L}}{\bf =q\times }\widehat{{\bf p}}$ the orbital angular
momentum operator, are constants of motion: $\left[ H,\widehat{K}\right] =%
\widehat{0},\ \left[ H,\widehat{{\bf J}}\right] =\widehat{{\bf 0}}.$
Following a standard procedure \cite{IT}, the stationary states of energy $E$
can be written as 
\begin{equation}
\Psi _{E_{\nu }}({\bf q},t)=%
{\psi _{a}({\bf q,}t) \choose \psi _{b}({\bf q,}t)}%
=%
{\psi _{a}(q{\bf )}{\cal Y}_{jj_{3}l_{a}}(\widehat{{\bf q}}) \choose i\psi _{b}(q{\bf )}{\cal Y}_{jj_{3}l_{b}}(\widehat{{\bf q}})}%
\exp (-\frac{i}{\hbar }E_{\nu }t),  \label{f}
\end{equation}
where ${\cal Y}_{jj_{3}l}$ are the normalized total angular momentum
functions, with 
\begin{equation}
\widehat{{\bf L}}^{2}{\cal Y}_{jj_{3}l}=\hbar ^{2}l(l+1){\cal Y}%
_{jj_{3}l},\qquad \widehat{{\bf J}}^{2}{\cal Y}_{jj_{3}l}=\hbar ^{2}j(j+1)%
{\cal Y}_{jj_{3}l},\qquad \widehat{K}\ {\cal Y}_{jj_{3}l}=\hbar \kappa {\cal %
Y}_{jj_{3}l},  \label{ope}
\end{equation}
where $l_{a}=j\pm 1/2,$ $l_{b}=j\mp 1/2$ when $\kappa =\pm \left(
j+1/2\right) .$ Let us write $\widehat{{\rm g}}_{\nu }$ in the general form 
\begin{equation}
\widehat{{\rm g}}_{\nu }\equiv \frac{1}{2}{\rm g}_{\nu _{a}}\left( I+\beta
\right) +\frac{1}{2}{\rm g}_{\nu _{b}}\left( I-\beta \right) .
\end{equation}
where $\nu _{a,b}$ are labels defined in Eq.(\ref{gs}). Thus the Dirac
equation is equivalent to the set of first-order (nonlinear) differential
equations 
\begin{equation}
c{\bf \sigma \cdot }\widehat{{\bf p}}\psi _{b}({\bf q})=\left( E_{\nu
}-mc^{2}+{\rm g}_{\nu _{a}}\frac{Ze^{2}}{q}\right) \psi _{a}({\bf q}),\qquad
c{\bf \sigma \cdot }\widehat{{\bf p}}\psi _{a}({\bf q})=\left( E_{\nu
}+mc^{2}+{\rm g}_{\nu _{b}}\frac{Ze^{2}}{q}\right) \psi _{b}({\bf q}).
\label{p}
\end{equation}
From (\ref{f}), (\ref{ope}) and (\ref{p}) we find that 
\begin{eqnarray}
\left( \frac{d}{dr}R_{B}(r)-\frac{\kappa }{r}R_{B}(r)\right) &=&\left( \sqrt{%
\frac{M_{2}}{M_{1}}}-{\rm g}_{\nu _{a}}\frac{Z\alpha }{r}\right) R_{A}(r),
\label{de} \\
\left( \frac{d}{dr}R_{A}(r)+\frac{\kappa }{r}R_{A}(r)\right) &=&\left( \sqrt{%
\frac{M_{1}}{M_{2}}}+{\rm g}_{\nu _{b}}\frac{Z\alpha }{r}\right) R_{B}(r), 
\nonumber
\end{eqnarray}
where $R_{A,B}(r)\equiv r\psi _{a,b}(r)$ and 
\begin{equation}
r=\sqrt{M_{1}M_{2}}q,\qquad M_{1}=\frac{mc^{2}+E_{\nu }}{\hbar c},\qquad
M_{2}=\frac{mc^{2}-E_{\nu }}{\hbar c},\qquad Z\alpha =Z\frac{e^{2}}{\hbar c}%
\ .  \label{fa}
\end{equation}
Next we look for solutions in the form of series 
\begin{equation}
R_{A}(r)=\exp (-r)r^{s}\sum_{\mu =0}a_{\mu }r^{\mu },\qquad R_{B}(r)=\exp
(-r)r^{s}\sum_{\mu =0}b_{\mu }r^{\mu }.  \label{se}
\end{equation}
Thus from (\ref{de}) and (\ref{se}) we get 
\begin{equation}
\left\{ 
\begin{array}{l}
\left( s-\kappa \right) a_{0}-{\rm g}_{\nu _{b}}Z\alpha b_{0}=0, \\ 
{\rm g}_{\nu _{a}}Z\alpha a_{0}+\left( s+\kappa \right) b_{0}=0,
\end{array}
\right. \quad \quad \qquad \qquad \qquad \qquad \qquad \qquad \text{for }\mu
=0,  \label{co1}
\end{equation}
and 
\begin{equation}
\left\{ 
\begin{array}{l}
\left( s+\mu +\kappa \right) a_{\mu }-a_{\mu -1}-{\rm g}_{\nu _{b}}Z\alpha
b_{\mu }-\sqrt{M_{1}/M_{2}}b_{\mu -1}=0, \\ 
\left( s+\mu -\kappa \right) b_{\mu }-b_{\mu -1}+{\rm g}_{\nu _{a}}Z\alpha
a_{\mu }-\sqrt{M_{2}/M_{1}}a_{\mu -1}=0,
\end{array}
\right. \qquad \text{for }\mu >0.  \label{co2}
\end{equation}

Given that $a_{0},b_{0}\neq 0,$ from (\ref{co1}) we obtain 
\begin{equation}
s=\pm \sqrt{\left( \kappa ^{2}-\left( Z\alpha \right) ^{2}{\rm g}_{\nu _{a}}%
{\rm g}_{\nu _{b}}\right) }\backsimeq \pm \sqrt{\left( \kappa ^{2}-\left(
Z\alpha \right) ^{2}\right) }>-\frac{1}{2}\ .  \label{s}
\end{equation}
The negative sign must be excluded since it would make the functions $%
R_{A,B} $ singular at the origin$.$ Choosing $\mu =n^{\prime }+1$ and $%
a_{n^{\prime }+1}=b_{n^{\prime }+1}=0,$ in order to terminate the series, we
have that $b_{n^{\prime }}=-a_{n^{\prime }}\sqrt{M_{2}/M_{1}}$. Then from (%
\ref{co2}) we get 
\begin{equation}
2\left( s+n^{\prime }\right) \sqrt{M_{1}M_{2}}=Z\alpha \left( {\rm g}_{\nu
_{a}}M_{1}-{\rm g}_{\nu _{b}}M_{2}\right) .  \label{ee}
\end{equation}
Finally, from (\ref{s}) and (\ref{ee}) we obtain the energy eigenvalues from 
\begin{equation}
2\left( s+n-|\kappa |\right) \sqrt{\left( mc^{2}\right) ^{2}-E_{n\kappa }^{2}%
}=Z\alpha {\rm g}_{\nu _{a}}\left( mc^{2}+E_{n\kappa }\right) -Z\alpha {\rm g%
}_{\nu _{b}}\left( mc^{2}-E_{n\kappa }\right) \ ,  \label{E}
\end{equation}
where 
\begin{equation}
n\equiv n^{\prime }+|\kappa |=n^{\prime }+j+\frac{1}{2}
\end{equation}
is the{\it \ principal quantum number}. Note that for the point nucleus
there exist bound solutions $($for $\kappa =-1)$ only up to $Z\backsimeq
1/\alpha \sqrt{{\rm g}_{\nu _{a}}{\rm g}_{\nu _{b}}}\backsimeq 1/\alpha .$

Given the fact that $<1/q>_{\Psi _{n\kappa }^{^{{\rm (NR)}}}}\backsim
1/n^{2},$ we expect that $\left( {\rm g}_{\nu _{a}}-1\right) \varpropto 1/n,$
to be able to meet the factor $1/n^{3}$ in both (\ref{e}) and (\ref{e2}).
Thus, in the examples that follow, we shall choose a particular form for $%
{\rm g}_{\nu _{a}}$ and ${\rm g}_{\nu _{b}}$: 
\begin{eqnarray}
{\rm g}_{\nu _{a}} &\equiv &{\rm g}_{n\kappa l_{a}}\equiv 1-\lambda
_{n\kappa l_{a}}(\alpha ,Z\alpha )\left( 1-\left( <\widehat{H}_{D}>_{\Psi
_{n\kappa }^{{\rm (Dirac)}}}\right) ^{n}\right) ,  \label{gs} \\
{\rm g}_{\nu _{b}} &\equiv &{\rm g}_{n\kappa l_{b}}\equiv 1-\lambda
_{n\kappa l_{b}}(\alpha ,Z\alpha )\left( 1-\left( <\widehat{H}_{D}>_{\Psi
_{n\kappa }^{{\rm (Dirac)}}}\right) ^{n}\right) ,  \nonumber
\end{eqnarray}
where $\widehat{H}_{D}\equiv H_{D}/mc^{2}.$ The factors $\lambda _{n\kappa
l}(\alpha ,Z\alpha )$ may depend on the original Dirac-Coulomb potential
quantum numbers $\nu _{a,b}\equiv n\kappa l_{a,b\ },$ the binding (powers of 
$Z\alpha ),$ and radiative corrections (powers of $\alpha $). The term
containing the expectation value of $\widehat{H}_{D}$ has the required
property, namely 
\begin{equation}
\left( 1-\left( <\widehat{H}_{D}>_{\Psi _{n\kappa }^{{\rm (Dirac)}}}\right)
^{n}\right) =\frac{1}{2}\frac{1}{n}\left( Z\alpha \right) ^{2}+O\left(
\left( Z\alpha \right) ^{4}\right) \varpropto \frac{1}{n}.  \label{val}
\end{equation}
Thus the factors ${\rm g}_{n\kappa l}$ includes a nonrelativistic coupling 
{\it and} a radiative term: $\lambda _{n\kappa l}(\alpha ,Z\alpha )$ (see
Eq.(\ref{gs}))$,$ times a purely quantum relativistic factor: $\left(
1-\left( <\widehat{H}_{D}>_{\Psi _{n\kappa }^{{\rm (Dirac)}}}\right)
^{n}\right) .$

Now from (\ref{E}) and (\ref{gs}) we finally get 
\begin{equation}
2\left( s+n-|\kappa |\right) \sqrt{\left( mc^{2}\right) ^{2}-E_{n\kappa }^{2}%
}=2E_{n\kappa }Z\alpha {\rm g}_{n\kappa l_{a}}+O\left( mc^{2}\left( Z\alpha
\right) ^{3}\left( {\rm g}_{n\kappa l_{a,b}}-1\right) \right) .  \label{fe}
\end{equation}
On the right-hand side of (\ref{fe}) we are neglecting a term proportional
to $\left( Z\alpha \right) ^{3}\left( {\rm g}_{n\kappa l_{a,b}}-1\right) $
which is, for instance, of order $\alpha \left( Z\alpha \right) ^{5}$ for
the case of the Lamb-shift.

\subsection{The Lamb-shift}

As a first instance, we shall consider an approximation to the Lamb-shift.
To this end notice that $<H_{D}>_{\Psi _{n\kappa }^{^{{\rm (Dirac)}}}}\equiv
E_{n\kappa }(\widehat{{\rm g}}=I),$ with $H_{D}\equiv H(\widehat{{\rm g}}=I)$
the Dirac-Coulomb Hamiltonian, where the eigenstates $\Psi _{n\kappa }^{^{%
{\rm (Dirac)}}}$ satisfy the normalization condition 
\begin{equation}
\int_{{\Bbb R}^{3}}\Psi _{n\kappa }^{^{{\rm (Dirac)}}}({\bf q})\Psi
_{n\kappa }^{^{{\rm (Dirac)}}}({\bf q})\ d^{3}q=1.
\end{equation}
Let 
\begin{equation}
\lambda _{n\kappa l}(\alpha ,Z\alpha )\equiv \frac{8}{3\pi }\alpha \left(
\left( L_{nl}+\frac{19}{30}-2\ln \left( Z\alpha \right) \right) \delta
_{l,0}+\frac{3}{8}\frac{1}{\kappa \left( 2\left| \kappa \right| -1\right) }%
\left( 1-\delta _{l,0}\right) \right) .
\end{equation}
The {\it effective} gauge potential to be considered is of the form 
\begin{equation}
A_{0}^{{\rm (eff)}}(q)=\widehat{{\rm g}}_{n\kappa }^{{\rm (Lamb)}}\frac{%
Z\left| e\right| }{q},
\end{equation}
where 
\begin{equation}
{\rm g}_{n\kappa l}^{{\rm (Lamb)}}={\rm g}_{n\kappa l}^{{\rm (Lamb)}}(\alpha
,Z\alpha )=1-\lambda _{n\kappa l}(\alpha ,Z\alpha )\left( 1-\left( <\widehat{%
H}_{D}>_{\Psi _{n\kappa }^{{\rm (Dirac)}}}\right) ^{n}\right) .  \label{gl}
\end{equation}
Thus $\widehat{{\rm g}}_{n\kappa }^{{\rm (Lamb)}}$ diminishes the Coulomb
binding $-Z\alpha /q{\bf ,}$ and as a consequence, the $s$ levels are pushed
higher. From (\ref{E}) and (\ref{gl}) we find 
\begin{equation}
\left( s+n-|\kappa |\right) \sqrt{\left( mc^{2}\right) ^{2}-\left(
E_{n\kappa }^{{\rm (Lamb)}}\right) ^{2}}=Z\alpha {\rm g}_{n\kappa l_{a}}^{%
{\rm (Lamb)}}E_{n\kappa }^{{\rm (Lamb)}}+O\left( mc^{2}\alpha \left( Z\alpha
\right) ^{5}\right) .
\end{equation}
Expanding this equation in powers of $\alpha ,$ we find the spectrum 
\begin{equation}
\left( E_{n\kappa }^{{\rm (Lamb)}}-mc^{2}\right) /mc^{2}=-\frac{1}{2}\frac{1%
}{n^{2}}\left( Z\alpha \right) ^{2}+\left( \frac{3}{8}\frac{1}{n^{4}}-\frac{1%
}{2}\frac{1}{n^{3}}\frac{1}{\left| \kappa \right| }\right) \left( Z\alpha
\right) ^{4}  \label{elas}
\end{equation}
\[
+\frac{4}{3\pi }\frac{1}{n^{3}}\left( \left( L_{n\kappa }+\frac{19}{30}-2\ln
\left( Z\alpha \right) \right) \delta _{\kappa ,-1}+\frac{3}{8}\frac{1}{%
\kappa \left( 2\left| \kappa \right| -1\right) }\left( 1-\delta _{\kappa
,-1}\right) \right) \alpha \left( Z\alpha \right) ^{4}+O\left( \alpha \left(
Z\alpha \right) ^{5}\right) . 
\]
We observe that the last term in (\ref{elas}) corresponds to the Lamb-shift
splitting to the Dirac levels to order $\alpha (Z\alpha )^{4}$ \cite{LAU}$.$
The Coulomb potential for the hydrogen atom ($Z$ $=1$) and the corresponding
Coulomb-like {\it radiative} corrections for $n=4,8,12$, are shown in fig. 1.

\subsection{The hyperfine splitting}

The second instance regards the {\it hyperfine splitting} in the energy
levels of the hydrogen atom $(Z=1).$ In hydrogenic atoms, the interaction of
the magnetic moment of the orbital electron with the magnetic moment of the
nucleus leads to a splitting of the {\it fine} {\it structure} levels with
fixed orbital angular momentum $l$ $\left( \kappa =-1\right) $ into the {\it %
hyperfine structure} levels. In this case we choose 
\begin{equation}
{\rm g}_{n\kappa l}^{{\rm (Hyp)}}=1-\lambda _{\kappa l}\left( 1-\left( <%
\widehat{H}_{D}>_{_{\Psi _{n\kappa }^{{\rm (Dirac)}}}}\right) ^{n}\right) \ ,
\label{gh}
\end{equation}
where 
\begin{equation}
\lambda _{\kappa l}\equiv \frac{2}{3}g_{p}\left( \frac{m}{M_{p}}\right)
\left( \delta _{S,1}-3\delta _{S,0}\right) \delta _{l,0}\ ,
\end{equation}
with $g_{p}$ $=$ $2\left( 1+\kappa _{p}\right) ,$ $\kappa _{p}=1.79284,$ the 
$g$-factor of the neutron. Thus from (\ref{E}) and (\ref{gh}) we get 
\begin{equation}
2\left( s+n-|\kappa |\right) \sqrt{\left( mc^{2}\right) ^{2}-\left(
E_{n\kappa }^{{\rm (Hyp)}}\right) ^{2}}=2Z\alpha {\rm g}_{n\kappa l_{a}}^{%
{\rm (Hyp)}}E_{n\kappa }^{{\rm (Hyp)}}+O\left( mc^{2}\left( Z\alpha \right)
^{3}\right)
\end{equation}
from which we get

\begin{eqnarray}
\left( E_{n\kappa }^{{\rm (Hyp)}}-mc^{2}\right) /mc^{2} &=&-\frac{1}{2}\frac{%
\alpha ^{2}}{n^{2}}+\left( \frac{3}{8}\frac{1}{n^{4}}-\frac{1}{2}\frac{1}{%
n^{3}}\frac{1}{\left| \kappa \right| }\right) \alpha ^{4}  \label{ehy} \\
&&+\frac{1}{3}\frac{1}{n^{3}}g_{p}\left( \frac{m}{M_{p}}\right) \left(
\delta _{S,1}-3\delta _{S,0}\right) \delta _{\kappa ,-1}\alpha ^{4} 
\nonumber \\
&&+O\left( \alpha ^{6}\right) .  \nonumber
\end{eqnarray}
This is again the correct spectrum for the relativistic Dirac levels to
order $O\left( \alpha ^{4}\right) $ \cite{IT}.

Notice that the first term in both (\ref{elas}) and (\ref{ehy}) gives the
energy spectrum of the bound states in the non-relativistic approximation.
The second terms are the corresponding leading corrections to the Balmer
formula: the {\it fine splitting}. These expressions are a consequence of
the modification $Z\alpha \rightarrow \widehat{{\rm g}}_{n\kappa }Z\alpha $
introduced by the nonlinear factor $\left( <\widehat{H}>_{\Psi _{n\kappa }^{%
{\rm (Dirac)}}}\right) ^{n}$ $\left( \text{together with }\lambda _{n\kappa
}(\alpha ,Z\alpha )\text{ in the case of the Lamb-shift}\right) $, which
depend on the solution of the original $(\widehat{{\rm g}}_{n\kappa }=I)$
eigenvalue problem itself.

Finally, an interesting exercise to consider regards the eigenvalue problem
containing both corrections: the Lamb-shift and the {\it hyperfine splitting}
for the case of the hydrogen atom ($Z=1$). This is easily done by defining 
\begin{equation}
\widehat{{\rm g}}_{n\kappa }\equiv \widehat{{\rm g}}_{n\kappa }^{{\rm (Lamb)}%
}+\widehat{{\rm g}}_{n\kappa }^{{\rm (Hyp)}}-I.
\end{equation}
Replacing $\widehat{{\rm g}}_{n\kappa }$ in (\ref{fe}) and expanding in
powers of $\alpha $ yields 
\begin{eqnarray}
\left( E_{n\kappa }-mc^{2}\right) /mc^{2} &=&-\frac{1}{2}\frac{\alpha ^{2}}{%
n^{2}}+\left( \frac{3}{8}\frac{1}{n^{4}}-\frac{1}{2}\frac{1}{n^{3}}\frac{1}{%
\left| \kappa \right| }\right) \alpha ^{4}+\frac{1}{3}g_{p}\left( \frac{m}{%
M_{p}}\right) \frac{1}{n^{3}}\left( \delta _{S,1}-3\delta _{S,0}\right)
\delta _{\kappa ,-1}\alpha ^{4} \\
&&+\frac{4}{3\pi }\frac{1}{n^{3}}\left( \left( L_{n\kappa }+\frac{19}{30}%
-2\ln \left( \alpha \right) \right) \delta _{\kappa ,-1}+\frac{3}{8}\frac{1}{%
\kappa \left( 2\left| \kappa \right| -1\right) }\left( 1-\delta _{\kappa
,-1}\right) \right) \alpha ^{5}  \nonumber \\
&&+O\left( \alpha ^{6}\right) ,  \nonumber
\end{eqnarray}
which is the spectrum of the hydrogen atom to order $\alpha ^{5}.$ Notice
that any series expansion of $\ln \left( \alpha \right) $ rapidly overcomes
the relativistic (Dirac) corrections of $E_{n\kappa }$ beginning from order $%
\alpha ^{5}.$

\section{Higher order corrections}

To achieve a still better quantitative agreement, several contributions of
higher order must be included \cite{LAU}, namely corrections to order $%
mc^{2}\alpha \left( Z\alpha \right) ^{5}$, $mc^{2}\alpha \left( Z\alpha
\right) ^{6}$ and $mc^{2}\alpha ^{2}\left( Z\alpha \right) ^{4}.$ Here we
only want to examine the $mc^{2}\alpha \left( Z\alpha \right) ^{5}$ term
(the 2nd order {\it binding}) as the next correction to the Lamb-shift: $%
2s_{1/2}-2p_{1/2}$ (\ref{elas}) for the hydrogen atom (see fig. 2)$.$ To
this end we can approximately fit the corresponding QED theoretical value by
making the replacement 
\begin{equation}
\lambda _{n\kappa l}(\alpha ,Z\alpha )\rightarrow \left( 1+Z\alpha \right)
\lambda _{n\kappa l}(\alpha ,Z\alpha ),  \label{lam2}
\end{equation}
in the $\lambda _{n\kappa l}$ factors contained in (\ref{gl}). For $Z=1,$
the correction calculated with (\ref{E}), (\ref{gl}) and (\ref{lam2}), and
the QED value \cite{LAU}, both to order $\alpha \left( Z\alpha \right) ^{5}$%
, are respectively: 
\begin{equation}
\delta E\backsimeq 7.663\text{ MHz}\ \text{,\qquad }\left( \delta E\right) _{%
{\rm (QED)}}\backsimeq 7.243\text{ MHz\ ,}  \nonumber
\end{equation}
which corresponds to a $\backsim 5\%$ of discrepancy. These calculations
have been performed with the value $\alpha \backsimeq 1/137.036\backsimeq
7.\,297\,35\times 10^{-3}$ for the {\it fine structure constant.}

On the other hand, in fig. 2 we observe that there is good agreement between
the present approach and the QED numerical values for $Z=1,2,...,10.$ In
fact, the Lamb-shift for $Z=1,$ $n=2$, is given by 
\begin{equation}
\Delta E_{n=2}^{{\rm (Lamb)}}=1,046.54\text{ MHz\ ,\qquad }\left( \Delta
E_{n=2}^{{\rm (Lamb)}}\right) _{{\rm QED}}=1,046.45\text{ MHz\ ,}  \nonumber
\end{equation}
where 
\begin{equation}
\Delta E_{n=2}^{{\rm (Lamb)}}\equiv E^{{\rm (Lamb)}}\left( 2s_{1/2}\right)
-E^{{\rm (Lamb)}}(2p_{1/2}),
\end{equation}
(see the red line in fig. 2); while for $Z=10,$ $n=2,$ we get 
\begin{equation}
\Delta E_{n=2}^{{\rm (Lamb)}}=4,469.82\times 10^{3}\text{ MHz\ ,\qquad }%
\left( \Delta E_{n=2}^{{\rm (Lamb)}}\right) _{{\rm QED}}=4,860.51\times
10^{3}\text{ MHz\ ,}
\end{equation}
with a $\backsim 8$ \% of discrepancy.

For the {\it hyperfine structure} we make 
\begin{equation}
\lambda _{\kappa l}^{{\rm (Hyp)}}\rightarrow \lambda _{\kappa l}^{{\rm (Hyp)}%
}\left( \alpha ,Z\alpha \right) =\left( 1+\delta _{{\rm (Breit)}}+\delta
\left( \alpha ,Z\alpha \right) \right) \lambda _{\kappa l}^{{\rm (Hyp)}}\ ,
\end{equation}
where $\delta _{{\rm (Breit)}}=\left( 3/2\right) \left( Z\alpha \right)
^{2}, $ and $\delta \left( \alpha ,Z\alpha \right) $ contains higher order
corrections: radiative, binding, finite mass and structure of the proton 
\cite{CRE,BL}. Higher (radiative and binding) order corrections to the
Lamb-shift seem to be meaningless and difficult to reproduce by using this
simple approach .

\section{Outlook}

In this article we have introduced a {\it nonlinear} {\it effective}
Coulomb-like gauge potential (electromagnetic in nature) into the
Hamiltonian of a Dirac particle to describe the Lamb-shift and the {\it %
hyperfine structure} of the energy spectrum. In all the calculations we have
kept track of the corresponding orders in both the {\it fine structure} {\it %
constant} $\alpha $ and the combination $Z\alpha $ introduced by the $%
\widehat{{\rm g}}_{n\kappa }$ matrix factor. In Eq. (\ref{E}) there still is
room for further research. For instance, we can study systems (apart from
the example of the Lamb-shift) where $\widehat{{\rm g}}_{\nu }(\alpha )$ is
of order $\varsigma $ $>0$ ($\varsigma $ some integer number) in $\alpha $
and $Z\alpha .$ In addition to this, the prospect of defining an iterative
recursion procedure from (\ref{E}) and (\ref{gl}) in order to improve the
(eventually convergent) coupling constant $Z\alpha $ should be studied.

Much work needs to be done to give a precise characterization of the various
physical models included in Eq.(\ref{E}). Particularly, there still is
lacking a thorough explanation of the structure of the $\widehat{{\rm g}}%
_{\nu }(\alpha )$ factors contained in the different {\it effective}
Hamiltonians that we have considered.

This work was supported by Direcci\'{o}n de Investigaci\'{o}n, Universidad
de Concepci\'{o}n, through grant \#96.011.019-1.0, and Fondecyt through
grant \#1970995.

\newpage

{\LARGE Figure captions}

\bigskip

Fig. 1. Schematic plots of the Coulomb potential per unit charge $%
A_{0}(q)/\left| e\right| $ for the hydrogen atom and the corresponding
Coulomb-like (radiative) $\left| \lambda _{n\kappa l_{a}}(\alpha ,Z\alpha
)\right| A_{0}(q)/\left| e\right| $ ($\kappa =-1,$ $l_{a}=0$) corrections (%
{\it Rc) }for $n=4,8,12.$

\bigskip

Fig. 2. The lamb-shift $2s_{1/2}-2p_{1/2}$ for $Z=1,...,40.$ The
calculations with the present approach to order $\alpha \left( Z\alpha
\right) ^{4}$ (green color) are improved with the corrections to order $%
\alpha \left( Z\alpha \right) ^{5}$ (red color). The continuous lines are
drawn in order to visualize the corresponding patterns. There is good
agreement for values of $Z$ in the range $1,..,10$ with those of QED (see
Refs. \cite{LAU,MO}) with a {\it relative accuracy} between $0.01$ \%. ($%
Z=1) $ and $8$ \% ($Z=10).$

\end{document}